\documentclass{ifacconf}

\usepackage{natbib}            
\usepackage{graphicx}          

\begin{document}

\begin{frontmatter}

\title{Global stability of synchronous and out-of-phase oscillations in central pattern generators}


\author[First]{Alexandra S. Landsman} 
\author[Second]{Jean-Jacques Slotine}

\address[First]{ETH Zurich, Department of Physics, Zurich, Switzerland (email: landsmanster@gmail.com)}                                              
\address[Second]{MIT, Nonlinear Systems Lab, Boston, MA, USA (e-mail: jjs@mit.edu)}

\begin{abstract}
Coupled arrays of Andronov-Hopf oscillators are investigated.  These arrays can be diffusively or repulsively coupled, and can serve as central pattern generator models in animal locomotion and robotics.  It is shown that repulsive coupling generates out-of-phase oscillations, while diffusive coupling generates synchronous oscillations.  Specifically, symmetric solutions and their corresponding amplitudes are derived, and contraction analysis is used to prove global stability and convergence of oscillations to either symmetric out-of-phase or synchronous states, depending on the coupling constant.  Next, the two mechanisms are used jointly by coupling multiple arrays. The resulting dynamics is analyzed, in a model inspired by the CPG-motorneuron network that controls the heartbeat of a medicinal leech.  
\end{abstract}

\end{frontmatter}

\section{I. Introduction}

Central pattern generators (CPGs) are often modeled as coupled nonlinear
oscillators delivering phase-locked signals.  Some of their
applications include animal locomotion, robotics (see \cite{Seo:07},
\cite{Ijs:07}, and \cite{Ijs:08}), and other biological
rhythmic behaviors such as e.g. the generation of a
heartbeat, \cite{Buo:04}.  In this paper, we explore a system of
coupled Andronov-Hopf oscillators that can be used to model such
phenomena.  Repulsive coupling between the oscillators creates a
traveling wave that simulates salamander gate (\cite{Ijs:07}), and
with additional coupling architecture, the heartbeat of a leech
(\cite{Buo:04}).

Explicit rotational coupling between neighbouring oscillators has been
used in earlier work (see \cite{Pha:07}, \cite{Seo:07}) to generate a traveling wave
or out-of-phase state.  Here, we use inhibitory (repulsive) coupling
to achieve a similar effect in perhaps a more physical way, allowing
indeed the system itself to compute couplings achieving the
out-of-phase behavior.  While all our derivations aim at establishing
global convergence results, we found some significant qualitative
differences between these two types of couplings, such as the
existence of two degenerate out-of-phase states in repulsive coupling,
but not in rotational coupling.  In addition, while rotational
coupling preserves the amplitude of the coupled oscillators, repulsive
coupling increases the amplitude of oscillation.  In fact, this
increase in amplitude is the key factor behind generating high
frequency oscillation patterns in the heartbeat of a leech model, to
be analyzed in the paper.

The paper is organized as follows. In Part II, we obtain out-of-phase
and synchronous solutions for nearest neighbor coupled Andronov-Hopf
oscillators, and derive the condition for the existence of an
out-of-phase state.  In Part III, either the out-of-phase or the
synchronous solution is shown to be globally stable using techniques
from contraction theory (for original derivation of contraction theory 
see \cite{Loh:98}).  The stability of the two
different types of behavior is determined by the values of the
coupling constant, with a positive coupling constant generating the
out-of-phase state and a negative constant resulting in a globally
stable synchronous solution.  In Part IV, the synchronous and the
out-of-phase arrays are globally coupled to each other in a model
inspired by the CPG in the heartbeat of a leech.  Building on prior
results, the bifurcation value for the onset of high frequency
oscillations in the synchronous array is derived.

 

\section{II. Steady-state dynamics}
In this section we solve for the steady-state dynamics of the nearest neighbor coupled Andronov-Hopf oscillators, showing that both synchronous and out-phase solutions exist.  In the next section, we consider the global stability properties of these solutions applying techniques from contraction theory.  
We first consider a ring of nearest neighbor coupled Andronov-Hopf oscillators, of the form:  
\begin{equation}
{\bf{\dot{x}_j =  F (x_j)}} + k  \left(\bf{x_j - x_{j+1} - x_{j-1}} \right)
\label{eq:osc}
\end{equation}
where ${\bf{x_j}}=\{x_j, y_j\}$ is a two dimensional vector describing the dynamics of the $j$th oscillator, with 
$\bf{F(x)}$ given by:
\begin{equation}
\bf{F}\left(\begin{array}{c} x \\ y \end{array} \right)\ = \left(\begin{array}{c}
x-y-x^3 -x y^2 \\
x+y-y^3 - y x^2 \end{array} \right)\
\label{eq:fx}
\end{equation}
In complex form, Eqns. (\ref{eq:osc}) and (\ref{eq:fx}) can be expressed as:
\begin{equation}
\dot{z}_j = (\alpha + i \omega) z_j - |z_j|^2 z_j + k \left(z_j - z_{j-1} - z_{j+1} \right)
\label{eq:complex}
\end{equation}
where $z$ is a complex variable, given by: $z = x + i y$.  

\begin{figure}
\begin{center}
\includegraphics[height=7cm]{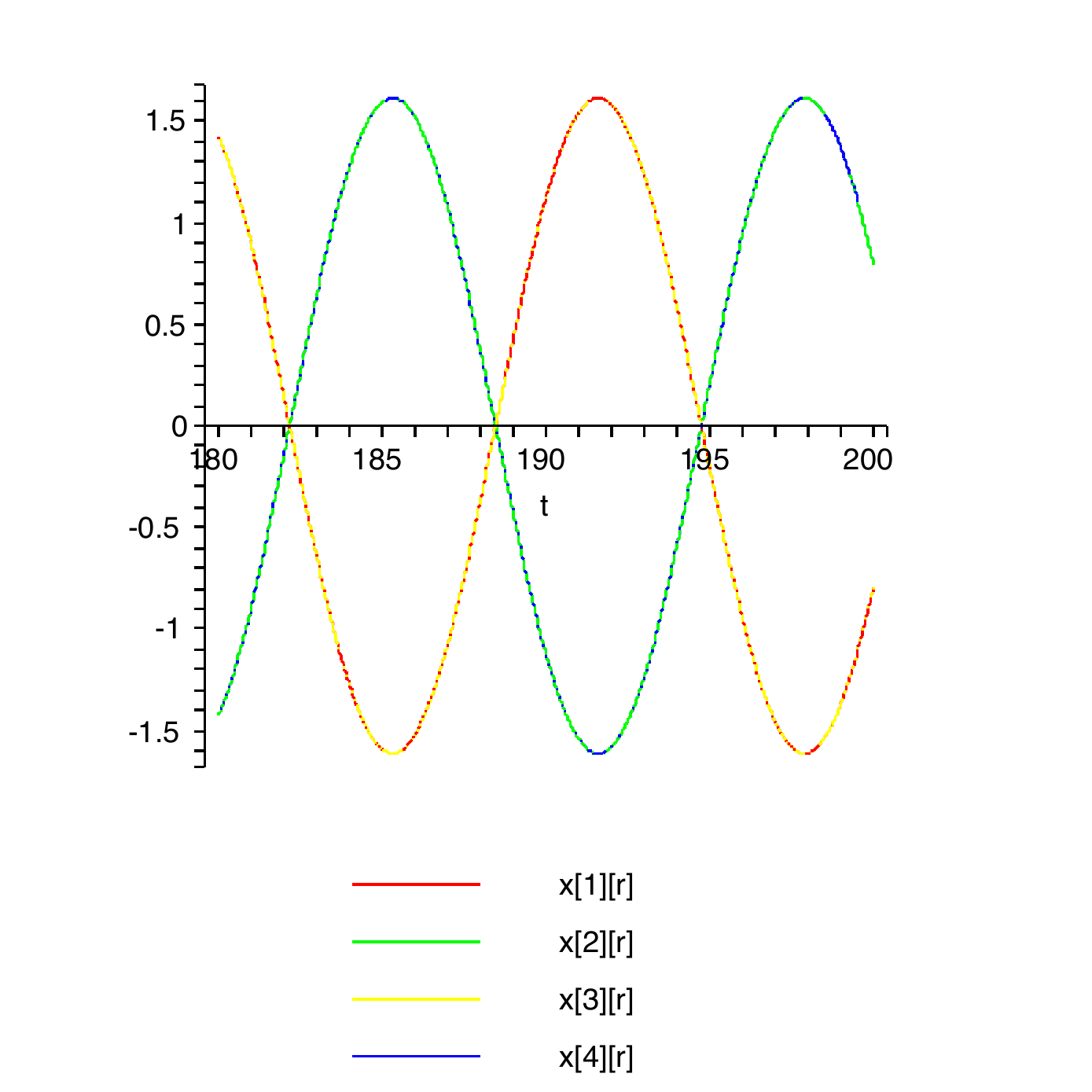}    
\caption{Repulsively coupled array, $k > 0$, for even number of oscillators,
 $N=4$.  Shows that the neighboring
  oscillators are maximally out-of-phase (by $\pi$)}  
\label{fig:2}                                 
\end{center}                                 
\end{figure}

In the absence of coupling, the dynamics are that of a limit cycle, of amplitude, $ |\bf{x_j}| = \sqrt{\alpha}$ and frequency $\omega$.  
Due to symmetry considerations, the synchronous state is one possible solution to the above equation, resulting in the amplitude of oscillation given by:  $ |{\bf{x_j}}| = \sqrt{\alpha - k}$.
This solution will be shown in the next section to be globally stable for diffusive type of coupling, given by: $k < 0$.  For the "repulsive coupling", given by $k > 0$, the system tends to an out-of-phase state, whereby the neighboring oscillators are maximally out of phase with each other, \cite{Lan:06}.  For oscillators coupled in a ring, this results in two different types of dynamics, depending on whether the number of oscillators, $N$ is even or odd.  When $N$ is even, the array oscillates with a difference of $\pi$ between nearest neighbors, splitting into two equal synchronous groups, that are $180$ degrees out of phase with each other (see Figure \ref{fig:2}, plotted for $N=4$).  
The phase difference between nearest neighbors is thereby given by:\begin{equation}
\triangle \phi_{j,j+1}^{even} = \pi
\label{eq:Neven}
\end{equation}
In the case of oscillators in a line, coupled with $k > 0$, the above phase difference is the only globally stable solution.  The situation becomes more
complicated in a ring coupled model when $N$ is odd.  In this case the phase difference of $\pi$
between nearest neighbors is not a symmetric or a stable solution.  
For example, imagine a ring of 3
oscillators, where the 2nd oscillator is out of phase with the 1st and the 3rd
by $\pi$.  Then the 1st and the 3rd oscillator will actually be in-phase,
which is an unstable state.  Requiring the neighboring oscillators to be maximally out of phase, while preserving the symmetry of the system leads to two possible degenerate solutions given by:     
\begin{equation}
\triangle \phi_{j,j+1}^{odd} = \pi \pm \pi/N
\label{eq:Nodd}
\end{equation}
The smallest phase difference between the two oscillators is given by the 
next to nearest neighbor phase difference:  $\triangle \phi_{j,j+2}^{odd} \pm 2 \pi/N$.  The
vectors $\bf{x_j}$ therefore fall on a circle where they are spaced with an equal phase difference of $2 \pi/N$, forming a symmetric out-of-phase solution.

\begin{figure}
\begin{center}
\includegraphics[height=7cm]{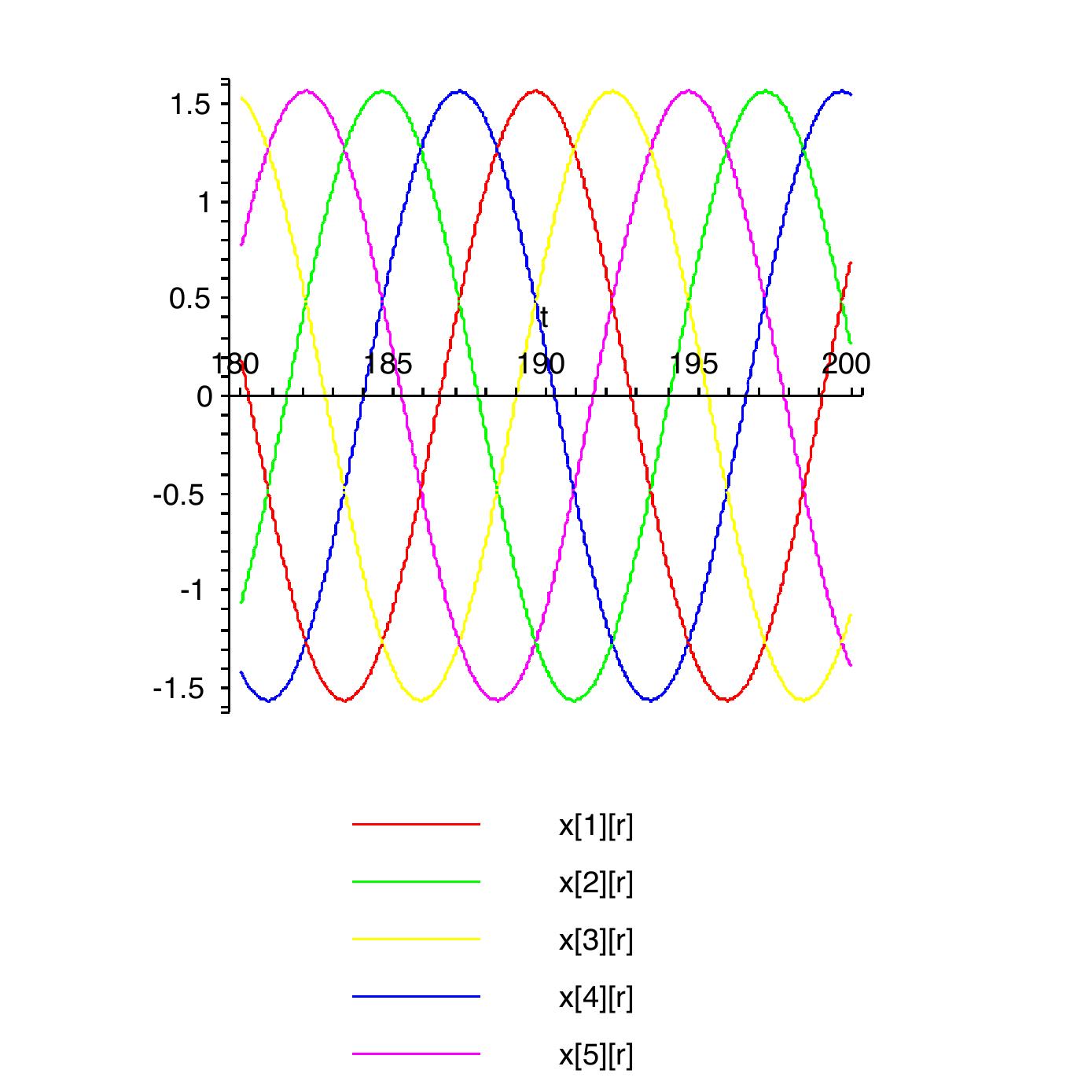}    
\caption{Repulsively coupled array for odd number of oscillators, 
$N=5$.  The neighboring oscillators are out of phase by $\pi+\pi/5$ (for
  example, compare
  the first and the second oscillator).}  
\label{fig:3}                                 
\end{center}                                 
\end{figure}

Figure \ref{fig:3} illustrates 
the stable out-of-phase dynamics for $N=5$.  Each oscillator in the
Figure is shifted in phase from its neighbor by $\pi + \pi/5$.
Comparing Eq. (\ref{eq:Nodd}) to
Eq. (\ref{eq:Neven}), we can see that as $N \rightarrow \infty$, the
steady-state dynamics of the $N$-odd  array 
approach that of $N$-even.  This should be expected, since for
 large $N$, adding one more oscillator to the ring (and thereby
changing $N$ from odd to even or vice versa) will
not significantly change the energy function, which is minimized when Eqs.
(\ref{eq:Nodd}) and (\ref{eq:Neven}) for $N$-odd and $N$-even, respectively,
are satisfied.
To solve for the amplitude of the out-of-phase solution, which is stable for repulsive coupling, we use Eqn. (\ref{eq:Nodd}), writing $\{z_{j-1} = e^{\pm i \left(\pi + \pi/N \right)} z_j, z_{j+1} = e^{\mp i \left(\pi + \pi/N \right)} z_j\}$ and substituting for $\{z_{j-1}, z_{j+1}\}$ into Eqn. (\ref{eq:complex}).  After grouping the linear terms, we can now solve for the amplitude of oscillation, getting:
\begin{equation}
|z_{j}|^{N-odd} 
= \lbrack \alpha + k \left( 1 + 2 cos(\pi/N)\right) \rbrack ^ {1/2}
\label{eq:AmpOdd}
\end{equation}
where the above equation is valid for all oscillators with odd $N$.
Performing the same type of analysis for even number of oscillators, where the nearest neighbors are $180$ degrees out of phase, we get:
\begin{equation}
|z_{j}|^{N-even} = \left(\alpha + 3 k \right)^{1/2}
\label{eq:AmpEven}
\end{equation}
Note that in the above equation the amplitude of the oscillation is independent of the total number of oscillators, while in Eqn. (\ref{eq:AmpOdd}), this amplitude increases with increasing $N$, asymptotically approaching $\left(\alpha + 3 k \right)^{1/2}$ as $n \rightarrow \infty$.
 
\section{III.  Contraction analysis}
This section uses partial contraction analysis 
to analyze the stability of the synchronous and out-of-phase states for the system in Eqns. (\ref{eq:osc}) - (\ref{eq:complex}), for the $N=3$ case.  
Here we use the partial contraction results first derived in the paper by Pham and Slotine, see \cite{Pha:07}.  The results state a simple sufficient condition for global exponential stability on a flow-invariant linear subspace $\mathcal{M}$ (i.e. a linear subspace $\mathcal{M}$ such that $\forall t:  {\bf{F}}(\mathcal{M}, t) \subset \mathcal{M}$) as given by:
\begin{equation}
- \lambda_{min} \left(\bf{V L V^T} \right) > sup \lambda_{max} \left(\frac{\partial \bf{F}}{\partial \bf{x}}\right)
\label{eq:condition}
\end{equation}
where $\bf{V}$ forms a basis of the linear subspace, $\mathcal{M^\perp}$ (orthogonal  to $\mathcal{M}$),  $\partial \bf{F}/\partial \bf{x}$ is the Jacobian of the uncoupled system, with $\bf{F}$
given by Eqn. (\ref{eq:fx}) and $\bf{L}$ is the coupling matrix, to be given below.  The terms $\lambda_{min}$ and $\lambda_{max}$ indicate the minimum and maximum eigenvalues of the symmetric parts of the matrices $\bf{L}$ and $\partial \bf{F}/\partial \bf{x}$, respectively.  Intuitively, the above condition insures that the system is contracting in the orthogonal subspace,  $\mathcal{M^\perp}$, thereby insuring that the dynamics converge exponentially to $\mathcal{M}$.

From Eqn. (\ref{eq:osc}), the coupling matrix, $\bf{L}$ for the $N=3$ case is given by:
\begin{equation}
\bf{L} = k
\left(\begin{array}{ccc}
I & -I & -I \\
-I & I & -I \\
-I & -I & I 
\end{array}\right)
\label{eq:something}
\end{equation} 
where $I$ above is a $2 \times 2$ identity matrix. 
The entire phase space, $\mathcal{M} \oplus \mathcal{M^\perp}$, is spanned by a total of six vectors:  the two vectors spanning the synchronous solution, plus the four vectors spanning a linear vector subspace formed by the two degenerate out-of-phase solutions.  For convenience, let's define the subspace corresponding to the synchronous case as:
\begin{equation}
\mathcal{M}_{sync} = \{\left(\bf{x, x, x} \right) : \bf{x} \in \mathcal{R}^2 \}
\label{eq:Msync}
\end{equation}
and the subspace corresponding to the out-of-phase states as:
\begin{equation}
\mathcal{M}_{phase} =\{ \left(\bf{x, R_{\frac{2 \pi}{3}} x,  R_{\frac{4 \pi}{3}} x} \right), 
  \left(\bf{x, R_{\frac{4 \pi}{3}} x,  R_{\frac{2 \pi}{3}} x} \right)  \}
\label{eq:Mphase}
\end{equation}
Where $\bf{R}$ is a $2 \times 2$ rotation matrix, with the rotation angles of $\{2 \pi/3, 4 \pi/3 \}$ obtained from Eqn. (\ref{eq:Nodd}) for the $N=3$ case.  The corresponding eigenvectors can be obtained by substituting ${\bf{x}} =\{1, 0 \}$ and $\{0, 1 \}$ into Eqns. (\ref{eq:Msync}) and 
(\ref{eq:Mphase}), resulting in six orthogonal eigenvectors.
In complex form the two out-of-phase solutions given in Eqn. (\ref{eq:Mphase}) can also be written as:  $z \cdot \left(1,  e^{i 2 \pi/3},  e^{i 4 \pi/3} \right)$ and $z \cdot \left(1,  e^{i 4 \pi/3},  e^{i 2 \pi/3} \right)$.  As before, the entire phase-space of solutions is spanned by the sum:  $\mathcal{M}_{sync}  \oplus \mathcal{M}_{phase}$.

Having obtained all the eigenvectors, we can now use Eqn. (\ref{eq:condition}) to analyze the stability of either the synchronous or the out-of-phase states, represented by $\mathcal{M}_{sync}$ and  $\mathcal{M}_{phase}$, respectively.  As will be shown shortly, this stability depends on the value of the coupling constant, $k$, with the synchronous solution being stable for negative $k$ (or diffusive type of of coupling) and the out-of-phase subspace being stable for positive values of $k$ (representing repulsive coupling).  

To analyze the stability of the synchronous state, we equate:  $\mathcal{M}_{sync} \equiv \mathcal{M}$ and $\mathcal{M}_{phase} \equiv \mathcal{M^\perp}$, thereby obtaining the $6 \times 4$ matrix $\bf{V}$ from Eqn. (\ref{eq:Mphase}).  The four eigenvalues of $ \bf{V L V^T}$, with $\bf{L}$ given in Eqn. (\ref{eq:something}), are all identical and given by:  $\lambda_{1,2,3,4} = 2 k$.  The eigenvalues of the symmetric part of the Jacobian,  $\partial \bf{F}/\partial \bf{x}$, are given by:  $\alpha - |x|^2$ and $\alpha - 3 |x|^2$, which are upper-bounded by $\alpha$.  It follows from Eqn. (\ref{eq:condition}) that the solution converges exponentially to  $\mathcal{M}_{sync}$ when,
\begin{equation}
\qquad k < -\alpha/2
\label{eq:synchstable}
\end{equation}
Next, doing the reverse by equating:  $\mathcal{M}_{phase} \equiv \mathcal{M}$ and $\mathcal{M}_{sync} \equiv \mathcal{M^\perp}$, we again obtain the eigenvalues of $ \bf{V L V^T}$ (with $\bf{V}$ now given by Eqn. (\ref{eq:Msync})):  $\lambda_{5,6} = -k $.  Combining the above results and again using Eqn. (\ref{eq:condition}), we now get the condition for the global stability of $\mathcal{M}_{phase}$:
\begin{equation}
k > \alpha
\label{eq:kstable}
\end{equation}
Equations (\ref{eq:synchstable}) and (\ref{eq:kstable}) give conditions for the global stability of synchronous and out-of-phase dynamics, respectively.  

It is instructive to compare the model in Eqns. (\ref{eq:osc}) and (\ref{eq:fx}) to another coupling architecture which directly uses rotational matrices, see \cite{Seo:07} and \cite{Pha:07}, to create globally stable out-of-phase state, in the following way (for a network of $N$ oscillators):
\begin{equation}
{\bf{\dot{x}_j = {\bf{F}} (x_j)}} + k  \left(\bf{x_j - R_{\frac{2 \pi}{N}} x_{j-1}}  \right)
\label{eq:oscR}
\end{equation}
The above model results in a globally stable state where the nearest neighbors are displaced out-of-phase by $2 \pi/N$.  Unlike the model in Eqn. (\ref{eq:oscR}), the coupling term in  Eqn. (\ref{eq:osc}) does not need to be adjusted to get the $2 \pi/N$ out-of-phase state as more oscillators are added, provided that the number, $N$, is always increased by 2, so that $N$ remains odd.  As described in the previous section, for $N$-even, the system splits into two identical synchronous groups $180$ degrees out-of-phase with each other.

The other significant differences of the out-of-phase state created by the coupling term in Eqn. (\ref{eq:oscR}) from the system analyzed in the present work include:   I. Existence of out-of-phase solution for any number, $N$, of oscillators.  This is in contrast to the system analyzed here, where (as mentioned above), an odd $N$ is needed for an out-of-phase state,  II. The existence of a single (non-degenerate) out-of-phase solution, unlike the two solutions given in Eqn. (\ref{eq:Mphase}),  III. Different phase difference between nearest neighbors.  Thus, the phase difference given by Eqn. (\ref{eq:Nodd}) is such that the nearest neighbors are maximally out-of-phase, while the phase difference from rotational coupling is such that oscillator $j$ is advanced from $j-1$ by a phase of $2 \pi/N$, and IV. Preservation of uncoupled amplitude, $\sqrt{\alpha}$ in the out-of-phase state.  This can be seen directly by substituting the out-of-phase solution $\bf{x_j = R_{\frac{2 \pi}{N}} x_{j-1}}$ into Eqn. (\ref{eq:oscR}), whereby the coupling term drops out, 

Points I-III can be explained by pointing out that in Eqn. (\ref{eq:oscR}), the rotational coupling creates an out-of-phase state in the same way that a synchronous state is  created in diffusive coupling.  In other words, the oscillators all try to be in synch with the rotated by $2 \pi/N$ solution of their nearest neighbor.  Therefore the out-of-phase state in rotational coupling  is actually analogous to the synchronous state in diffusive coupling, which also has uniqueness, global stability (for appropriate values of $k$), and existence for any value of $N$.   Point IV, that is the increase in the amplitude of oscillation (given by Eqn. (\ref{eq:AmpOdd})) in the out-of-phase state, is actually essential for the creation of $N \omega$ frequency  oscillations when the two arrays are globally coupled.  This model, inspired by the nervous system of a leech is analyzed in the following section.  

 
\section{IV. Globally coupled arrays and the onset of high frequency oscillations}

Here we globally couple two arrays, each described by Eqns. (\ref{eq:osc}) and (\ref{eq:fx}), with the only difference being that the first array has repulsive coupling, given by $k_r>0$ and the second array has diffusive coupling, given by $k_d<0$.  The dynamics were inspired by the CPG-motorneuron network that controls the heartbeat of a medicinal leech, \cite{Buo:04}.  The heartbeat of the leech is driven by direct contact between two arrays of motorneurons, such that on one side of the leech the heart beats in a rear-to-front (peristaltic) fashion, well described by the out-of-phase state of coupled limit cycle oscillators.  On the other side, the heart beats synchronously and is therefore represented by the diffusively coupled array, where the synchronous state is stable.  The total system has the following form:
\begin{equation}
{\bf{\dot{x^r}_j = F (x^r_j)}} + k_r  \left({\bf{x^r_j - x^r_{j+1} - x^r_{j-1} }}\right) + c  \sum_{k=1}^N  |\bf{x^d_k}|
\label{eq:osc2}
\end{equation}
\begin{equation}
{\bf{\dot{x^d_j} = {\bf{F}} (x^d_j)}} + k_d  \left({\bf{x^d_j - x^d_{j+1} - x^d_{j-1} }}\right) +  c \sum_{k=1}^N  |\bf{x^r_k}| 
\label{eq:osc3}
\end{equation}
where as before, $\bf{F(x)}$ is defined in Eqn. (\ref{eq:fx}).
Based on the results of the previous section, we know that the system in Eqn. (\ref{eq:osc2}) has a stable out-of-phase state (corresponding to peristaltic motion), while the system in Eqn. (\ref{eq:osc3}) has a stable synchronous state.  

\begin{figure}
\begin{center}
\includegraphics[height=7cm]{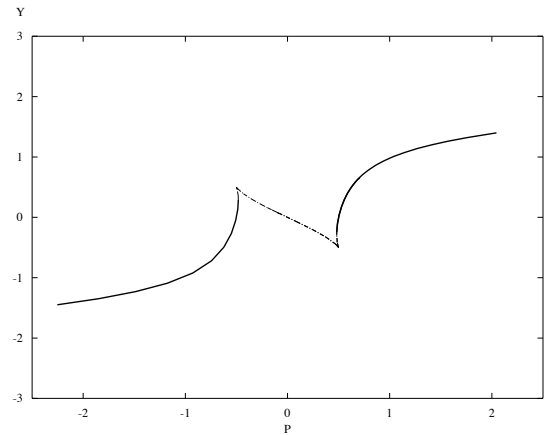}    
\caption{Bifurcation diagram as a
  function of $P$, ($\alpha=0.9$, $k_d= -0.1$, $\omega=1/2$).  Bold lines correspond to the high-frequency oscillations
  after the system in Eqn. (\ref{eq:complex2}) has been driven through a bifurcation for 
$P>P_{bif}$.}  
\end{center} 
\label{fig:3b}                                
\end{figure}

It has been shown both computationally (see \cite{Pal:05}) and analytically (see \cite{Lan:06}) that a system of this type undergoes a bifurcation, as the global coupling constant, $c$, increases.  For $c > c_{bif}$, the $\{\bf{x^d}\}$ array begins to oscillate in phase at the ultra-harmonic frequency given by $N \omega$.    Following the method in Landsman and Schwartz, \cite{Lan:06}, the bifurcation value of $c$ that leads to high frequency oscillations can be calculated by solving for the value of the parameter $P$ that causes a Hopf bifurcation in the following equation:
\begin{equation}
\dot{z}_j = (\alpha - k_d + i \omega) z_j - |z_j|^2 z_j + P
\label{eq:complex2}
\end{equation}
where we  have used the complex formulation, $z=x + i y$, of Eqn. (\ref{eq:complex}) for the diffusively coupled array and substituted the synchronous solution: $z_j = z_{j+1} = z_{j-1}$. 
The bifurcation diagram for Eqn. (\ref{eq:complex2}) as a function of $P$ is plotted in Figure (\ref{fig:3b}), where bold lines at higher $|P|$ indicate stable equilibria, with the broken line in the center showing an unstable equilibria.  

\begin{figure}
\begin{center}
\includegraphics[height=7cm]{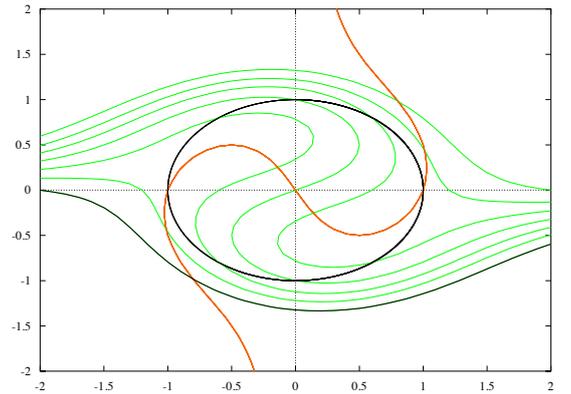}    
\caption{Nullclines for Eqn. (\ref{eq:complex2}) for different values of $P$ ($\alpha=0.9$, $k_d= -0.1$, $\omega=1/2$).
The green nullclines are the $\dot{x}=0$ nullclines for
various values of $P$.  The red nullcline is the $\dot{y}=0$
nullcline.  Higher nullclines corresponding to higher values
of $P$.   The circular orbit is a limit cycle for $P =0$.}  
\label{fig:4}                                 
\end{center}                                 
\end{figure}  

The bifurcation diagram in Fig. (\ref{fig:3b}), in fact also corresponds to the bifurcation digram for the onset on high frequency oscillations in the diffusively coupled array, given by Eqn. (\ref{eq:osc3}).  Namely the solid lines in the figure correspond to the $N \omega$ frequency oscillations in the $\bf{x^d}$ array, while the broken line corresponds to oscillations close to the limit cycle frequency, $\omega$ in the same array.  We can therefore obtain the bifurcation value of the global coupling, $c$ for the onset of $N \omega$ frequency oscillations by first solving for the bifurcation value of $P$ (given by $P_{bif}$), and then using:  $P_{bif} \approx c _{bif} N |z^r|$ to calculate $c_{bif}$.  Here $|z^r|$ is the amplitude of the out-of-phase oscillation in the repulsively coupled array.  It is given by Eqn.   (\ref{eq:AmpOdd}), with $k \rightarrow k_r$.  Solving for $c_{bif}$, we obtain the value of the bifurcation to $N \omega$ oscillations as a function of $P$:
\begin{equation}
c_{bif} \approx \frac{P_{bif}}{N |z^r|} = \frac{P_{bif}}{N  \lbrack \alpha + k_r \left( 1 + 2 cos(\pi/N)\right) \rbrack ^ {1/2}}
\label{eq:general}
\end{equation}
where  Eqn. (\ref{eq:AmpOdd}) was used in the denominator. 

The effect of the constant $P$ on the dynamics of Eqn. (\ref{eq:complex2}) can be seen by referring to the nullclines diagram, in  Figure \ref{fig:4}.  The various green curves correspond to the $\dot{x} =0$ nullclines plotted for the corresponding values of $P$, with the $P=0$ nullcline crossing the origin.  Since $P$ is a real constant in Eqn. (\ref{eq:complex2}), the $\dot{y}=0$ nullcline (shown in red) does not change with the parameter $P$.  As $|P|$ increases, the system bifurcates to a steady-state with a stable fixed point given by the intersection of the corresponding nullclines.  The bifurcation value of $|P|$ is found at a point where the intersection of the vertical line with the $\dot{x}=0$ nullcline no longer has three real roots (see for example \cite{Guck:83}), and given by:
\begin{equation}
P_{bif} =  \frac{\left(8 \gamma^2 + \omega^2 \right)}{4 \gamma^{1/2}}
\label{eq:Cbnew}
\end{equation}
where $\gamma = \left(\alpha - k_d \right)/3$.  Note that $\gamma > 0$, since $k_d < 0$ in diffusive coupling.  
Substituting Eqn. (\ref{eq:Cbnew}) into Eqn. (\ref{eq:general}), we have the bifurcation value as a function of $\alpha$, $N$, $\omega$, and the repulsive and diffusive coupling constants, given by $k_r$ and $k_d$, respectively.  

The mechanism behind the onset of ultraharmonics for a similar type of coupling architecture was analyzed in \cite{Lan:06}.  Here we briefly summarize the mechanism behind the onset.  The onset of high frequency oscillations hinges on the amplitude of the repulsively coupled array being higher than the amplitude of the diffusively coupled array.  This results in a diffusively coupled array being driven through a bifurcation first, when $c > c_{bif}$, with $c_{bif}$ given in Eqn. (\ref{eq:general}), and thereafter being driven by the out-of-phase dynamics of the repulsively coupled array, which causes the high-frequency synchronous oscillation in ${\bf{x^d}}$.   In the language of contraction theory, \cite{Loh:98}, for $c > c_{bif}$, the diffusively coupled array becomes a contracting system and can therefore be driven at the ultraharmonic frequency provided by the repulsively coupled array, which after the bifurcation acts like a drive.  If $c$ is increased even higher, beyond the bifurcation value of the repulsively coupled array, then total oscillator death in both arrays results.  For $c$ below this critical value, but above $c_{bif}$, given by Eqns. (\ref{eq:general}) (with $P_{bif}$ given by Eqn. (\ref{eq:Cbnew})), high frequency synchronous oscillations are produced in the diffusively coupled array.

\end{document}